# Acoustic Echo Cancellation Postfilter Design Issues For Speech Recognition System

Urmila Shrawankar
IEEE Student Member & Research Scholar, (CSE),
G H Raisoni College of Engg., Nagpur, INDIA
urmila@ieee.org

Dr. Vilas Thakare
Professor & Head, PG Dept. of Computer Science,
SGB Amravati University, Amravati, INDIA

*Abstract:*

**In this paper a generalized postfilter algorithm design issues are presented. This postfilter is used to jointly suppress late reverberation, residual echo, and background noise. When residual echo and noise are suppressed, the best result obtains by suppressing both interferences together after the Acoustic echo cancellation (AEC).**

**The main advantage of this approach is that the residual echo and noise suppression does not suffer from the existence of a strong acoustic echo component.**

**Furthermore, the Acoustic echo cancellation (AEC) does not suffer from the time-varying noise suppression.**

**A disadvantage is that the input signal of the Acoustic echo cancellation (AEC) has a low signal-to-noise ratio (SNR). To overcome this problem, algorithms have been proposed where, apart from the joint suppression, a noise-reduced signal is used to adapt the echo canceller.**

## I. INTRODUCTION

The speech signals collected from microphone do not only contain the desired near-end clean speech but also interferences such as room reverberation that is caused by the near-end source, background noise and a far-end echo signal that results from the acoustic coupling between the loudspeaker and the microphone. These interferences degrade the fidelity and intelligibility of near-end speech.

Dereverberation via suppression and enhancement is similar to noise suppression. Techniques have been prepared [14] to reduce reverberation effects. One approach is to remove reverberation effects is by passing the reverberant signal through a second filter that inverts the reverberation process and recover the original signal.

Many efforts to improve adaptive filters for non-linear environments [22] have been reported and two dominant solutions have emerged.

1) The Volterra filter : The Volterra [9] solution is generally slow to convergence and is highly computationally complex.

2) The post-filtering in combination with AEC adaptive filtering. : Post-filters are less complex but rely on the performance of linear adaptive filters that are still disturbed by non-linear echo.

Recently, feasible spectral enhancement techniques to suppress reverberation have emerged. A spectral variance estimator for the late residual echo that results from the deficient length of the adaptive filter is derived. Both estimators are based on a statistical reverberation model. The model parameters depend on the reverberation time of the room, which can be obtained using the estimated acoustic echo path.

A frame work of a postfilter is discussed which suppresses late reverberation of the near-end speech, residual echo and background noise, and maintains a constant residual background noise level.

Acoustic echo cancellation is the most important and well-known technique to cancel the acoustic echo [4,7]. The acoustic echo cancellation problem is usually solved by using an adaptive filter in parallel to the acoustic echo path [8,10]. The adaptive filter is used to generate a signal that is a replica of the acoustic echo signal. An estimate of the near-end speech signal is then obtained by subtracting the estimated acoustic echo signal, i.e., the output of the adaptive filter, from the microphone signal.

In practice, there is always residual echo, i.e., echo that is not suppressed by the echo cancellation system. The residual echo results from

1) The deficient length of the adaptive filter,

2) The mismatch between the true and the estimated echo path, and

3) Nonlinear signal components.

Wiener filter is used to suppress the echo and noise simultaneously. A postfilter which follows the traditional single microphone acoustic echo canceller (AEC).

The postfilter jointly suppresses reverberation of the near-end speaker, residual echo, and background noise. The near-end speech signal is estimated using an optimally-modified log spectral amplitude (OM-LSA) estimator which requires an estimate of the spectral variance of each interference. Here, a modified version of the OM-LSA estimator [23] is considered to obtain an estimate of the spectral component. The OM-LSA spectral gain function, which minimizes the mean-square error of the log-spectra. Different estimators can be used to estimate the a priori signal-to-interference ratio (SIR), e.g., the decision direct estimator developed by Ephraim and Malah [24] or the recursive causal or non causal estimators developed by Cohen [25]. In the sequel, the decision directed estimator is used for the estimation of the a priori SIR. A lower-bound on the a priori SIR that helps to reduce the amount of musical noise. The weighting factor controls the tradeoff between the amount of noise reduction and transient distortion introduced into the signal.

The weighting factor is commonly close to a larger value of results in a greater reduction of musical noise.

In addition, we discuss the estimation of the a priori signal-to-interference ratio (SIR), which is necessary for the OM-LSA estimator. The late residual echo and the late reverberation spectral variance estimators require an estimate of the reverberation time and delay reductions with the help of FIR filter. A major advantage of the AEC scenario is that due to the existence of the echo an estimate of the reverberation time can be obtained from the estimated acoustic echo path.

The delay reductions can be achieved by filtering degraded speech signals in the time domain with finite impulse response (FIR) filters [21].

Approaches to calculate such an FIR filter include

1) Filter Bank Equalizer (FBE),
2) Low Delay Filter (LDF),
3) Inverse Discrete Fourier Transform (IDFT) of spectral gains.

The rest of the paper is arranged as in sec II a postfilter algorithm designing is presented. In sec III Issues related to Performance Evaluation of the algorithms is discussed, in sec IV some echo cancellers and their comparative study is discussed. Finally conclusion is given in the V section.

## II. ALGORITHM OUTLINE AND DISCUSSION

The steps for a complete generalized algorithm designing of a postfilter that is used for the joint suppression of residual echo, late reverberation, and back-ground noise. This postfilter is used in conjunction with a standard Acoustic echo cancellation (AEC), that includes,

1) The Acoustic Echo Path Estimation
2) The Estimation of the Spectral Variance of the Interferences
3) The OM-LSA Gain Function

*A. Generalized Postfilter Algorithm Outline*

The Major Steps are [1]:

1) Acoustic Echo Cancellation

2) Estimate Reverberation Time

3) STFT (Short-Time Fourier Transform)

4) Estimate Background Noise

5) Estimate Late Residual Echo Spectral Variance

6) Estimate Late Reverberant Spectral Variance

7) Postfilter:

   a) Calculate the *a posteriori*

   b) Calculate the speech presence probability

   c) Calculate the gain function

   d) Calculate the spectral speech component

8) Inverse STFT (Short-Time Fourier Transform) : Calculate the output by applying the inverse short-time fourier transform (STFT) to The spectral speech component

*B. The Acoustic Echo Path Estimation*

A standard normalized least mean square (NLMS) algorithm is used to estimate part of the acoustic echo path. Some other, algorithms are also available, e.g., Recursive Least Squares (RLS) or Affine Projection (AP) [6]. Since it is sparse in nature, the improved proportionate NLMS (IPNLMS) algorithm can be used, proposed by Benesty and Gay [16].

In general, the residual echo signal is not zero because of the deficient length of the adaptive filter, the system mismatch and nonlinear signal components that cannot be modeled by the linear adaptive filter. While many residual echo suppressions [11,13] focus on the residual echo that results from the system mismatch.

Double-talk occurs during periods when the far-end speaker and the near-end speaker are talking simultaneously and can seriously affect the convergence and tracking ability of the adaptive filter. Double-talk detectors and optimal step-size control methods have been presented to ease this problem [10,11,17,18].

The ultimate goal is to obtain an estimate of the anechoic speech signal. While the Acoustic echo cancellation (AEC) estimates and subtracts the far-end echo signal a postfilter is used to suppress the residual echo and background noise. The postfilter is usually designed to estimate the reverberant speech signal or the noisy reverberant speech signal.

The reverberant speech signal can be divided into two components:

1) The early speech component, which consists of a direct sound and early reverberation that is caused by early reflections, and

2) The late reverberant speech component, which consists of late reverberation that is caused by the reflections that arrive after the early reflections, i.e., late reflections.

Independent research [15,19,20], has shown that the speech quality and intelligibly are most affected by late reverberation.

### C. The Estimation Of The Spectral Variance Of The Interferences

*i. Late Residual Echo Spectral Variance Estimation*

Enzner [12] proposed a recursive estimator for the short-term power spectral density (PSD) of the late residual echo. The recursive estimator exploits the fact that the exponential decay rate of the acoustic impulse response (AIR) is directly related to the reverberation time of the room, which can be estimated using the estimated echo path. Additionally, the recursive estimator requires a second parameter that specifies the initial power of the late residual echo.

Furthermore, in many applications, the distance between the loudspeaker and the microphone is small, which results in a strong direct echo. The presence of a strong direct echo results in an erroneous estimate of both the reverberation time and the initial power [27].

*ii. Late Reverberant Spectral Variance Estimation*

It requires an estimator for the late reverberant spectral variance of the near-end speech signal. The parameter controls the time instance (measured with respect to the arrival time of the direct sound).

### D. The OM-LSA GAIN Function

When the early speech component is assumed to be active, the log spectral amplitude (LSA) gain function is used. Under the assumption that early speech component and the interference signals are mutually uncorrelated, The OM-LSA spectral gain function, which minimizes the mean-square error of the log-spectra, is obtained as a weighted geometric mean of the hypothetical gains associated with the speech presence probability denoted [26].

### E. Short-Time Fourier Transform (STFT)

Postfilters that are capable of handling both the residual echo and background noise are often implemented in the STFT [5] domain.

In general, they require two STFT and one inverse STFT, which is equal to the number of STFTs used in the proposed solution. The computational complexity of the proposed solution is comparable to former solutions since the estimation of the reverberation time and the late reverberant spectral variance only requires a few operations. The computational complexity of the Acoustic echo cancellation (AEC) [4] can be reduced by using an efficient implementation of the AEC in the frequency domain [28], rather than in the time-domain.

### III. PERFORMANCE EVALUATION OF THE ALGORITHM

The ability of the postfilter to suppress background noise and non-stationary interferences, i.e., late residual echo and late reverberation is depend on,

1) Residual Echo Suppression
2) De-reverberation Performance Evaluation
3) Joint Suppression Performance

Explained in further sections.

### A. Residual Echo Suppression [39]

The performance of the late residual echo spectral variance estimator and its robustness with respect to changes in the tail of the acoustic echo path.

The echo cancellation performance, and the improvement due to the postfilter, can be estimated using the echo return loss enhancement (ERLE). A small amount of residual echo may be audible in the processed signal. However, in the presence of background noise, the residual echo in the processed signal is masked by the residual noise.

The robustness of the late residual echo suppressor with respect to changes in the tail of the acoustic echo path when the far-end speech signal is active.

Since the late residual echo estimator is mainly based on the exponential decaying envelope of the acoustic impulse response (AIR), which does not change over time, the postfilter does not require any convergence time and it does not suffer from the change in the tail of the acoustic echo path. Furthermore, during double-talk, the adaptive filter might not be able to converge due to the low echo to near-end speech-plus-noise ratio of the microphone signal.

### B. De-reverberation Performance Evaluation

The de-reverberation performance of the near-end speech in the presence of background noise.

The de-reverberation performance can be evaluated using the segmental SIR and the log spectral distance (LSD).

The segmental signal-to-noise ratio (SNR) value can be calculated by averaging the instantaneous SNR of those frames where the near-end speech is active. Since the non-stationary interferences, such as the late residual echo and reverberation, are suppressed down to the residual background noise level the postfilter will always include the noise suppression. The performance of the dereverberation process depend on,

1) The segmental SIR and LSD measures for the unprocessed signal,
2) The processed signal [noise suppression (NS) only]

3) The processed signal without direct path compensation DPC [noise and reverberation suppression (NS+RS)],
4) The processed signal with direct path compensation DPC (NS+RS+DPC).

C. *Joint Suppression Performance*

The performance of the entire system when all interferences are present, like, during double-talk.

The performance of the entire system during double-talk can be evaluated using the segmental SIR and the LSD at three different segmental SNR values. The suppression of each additional interference results in an improvement of the performance. Since all non-stationary interferences, i.e., the late residual echo and reverberation, are reduced down to the residual background noise level, the background noise is to be suppressed first.

The Performance evaluation can be done using

1) Acoustic echo cancellation (AEC),
2) AEC and postfilter (noise suppression),
3) AEC and postfilter (noise and residual echo suppression), and
4) AEC and postfilter (noise, residual echo, and reverberation suppression)

## IV. ECHO CANCELLATION ALGORITHMS & COMPARETIVE STUDY

Linear adaptive filtering [43] is still popular and it is of interest to assess adaptive filters in such environments. Traditionally, adaptive filters have been deployed to achieve AEC by estimating the acoustic echo response using algorithms [2,3] such as Echo Return Loss Enhancement (ERLE), Least Mean Square (LMS), Normalized-LMS (NLMS), Affine Projection Algorithm (APA), Frequency Block-LMS (FBLMS)) algorithm [14]. Several approaches have been proposed over recent years to improve the performance of the standard NLMS algorithm in various ways for AEC. These include Fourier [30] and wavelet [31] based adaptive algorithms, variable step-size (VSS) algorithms [32,33], data reusing techniques [34,35], partial update adaptive filtering techniques [36,37] and subband adaptive filtering (SAF) schemes [38]. These approaches aim to address issues in echo cancellation including the performance with colored input signals, time-varying echo paths and computational complexity. In contrast to these approaches, sparse adaptive algorithms have been developed specifically to address the performance of adaptive filters in sparse system identification.

In this section we have discussed some AEC algorithms [40,22] followed by comparative study with respective to convergence and performance.

1. ERLE: Echo Return Loss Enhancement, The adaptive filters performance is measure by comparing the degradation in echo return loss enhancement ERLE with linear echo and linear & non-linear echo. The traditional ERLE calculated after removing the near-end source signal so that the true amount of echo cancellation can be calculated during noisy time.

2. LMS : Least Mean Square algorithm [32,33,34] is most commonly used adaptive algorithm. It is simple and gives reasonable performance. Since it is an iterative algorithm it can be used in a highly time-varying signal environment. It has a stable and robust performance against different signal conditions. It converges with slow speeds when the environment yields a correlation matrix R possessing a large eigenspread.

3. FBLMS: Frequency Block-LMS [22] is an implementation of a block-by-block LMS using fast convolution. In block LMS, the input signal is divided into blocks and weights are updated block wise. For long adaptation processes the Block LMS is used to make the LMS faster.

4. NLMS: The Normalized Least-Mean-Square algorithm [38] is one of the most popular for AEC due to its straightforward implementation and low complexity compared to, other algorithms, for example, the Recursive Least Squares (RLS) algorithm. The Normalized LMS (NLMS) introduces a variable adaptation rate. It improves the convergence speed in a non-static environment.

5. PNLMS: The PNLMS (proportionate NLMS) [3,16,29,41,42,44] and MPNLMS (μ PNLMS ) [42] algorithms have been proposed for sparse system identification. It prevents the coefficients from stalling when they are much smaller than the largest coefficient.

6. MPNLMS ( μ PNLMS) algorithm [31] is proposed to improve the convergence of PNLMS. MPNLMS converges fast since it allows all filter coefficients to attain a converged value to within a vicinity of their optimal value in the same number of iterations from the initial iteration.

7. IPNLMS: The IPNLMS (Improved PNLMS) [41] algorithm was originally developed for NEC and was further developed for the identification of acoustic room impulse responses. It employs a combination of proportionate (PNLMS) and non-proportionate (NLMS) adaptation. IPNLMS [44] is a very good approximation of the filter, while being more convenient from a practical point of view.

8. APA: Affine Projection Algorithm : Proportionate-type APAs [6,37,45] are attractive mainly for their fast convergence rate and tracking. However, these algorithms were derived based on a straightforward extension of PNLMS-type algorithms. The APA can be viewed as a generalization of the NLMS algorithm. Thus, the proportionate-type APAs were straightforwardly obtained from the proportionate-type NLMS algorithms.

An analysis on the comparative impact of additive noise and non-linear echo on the performance of adaptive filtering for linear acoustic echo cancellation (AEC) [21,40] is as follow

- ERLE convergence time and system distance metrics.
- ERLE is the mean ERLE obtained during a 10 second (50-60s) period where each algorithm has converged.
- NLMS & APA algorithms give similar performance
- The more computationally efficient FBLMS algorithm has adverse effect and gives poorer performance than the LMS.
- The FBLMS algorithm is the most affected. Performance decreases by about 90dB over the same range and for SNeRs less than 75dB performance is worse than that for the standard LMS algorithm.
- The LMS algorithm is the most robust of all adaptive filters considered; it has the least degradation in performance as the SNR or SNeR decreases.

A. *Performance of Algorithms in non-linear & noisy environments.*

- APA and NLMS algorithms show similar behavior in nonlinear environments
- APA and NLMS algorithms, give better performance in non-linear environments than noisy environments when the SNR < 100dB.
- FBLMS gives similar performance in nonlinear environments & noisy environments.
- APA and NLMS have comparable behaviour in nonlinear environments.
- The echo canceller seems to be more robust to non-linearities than noise with a similar SNR (with the exception of the FBLMS algorithm).
- The performance of APA, NLMS and FBLMS algorithms decreases by approximately the same amount in noisy environments
- FBLMS is badly affected in noisy environments,
- NLMS achieves approximately 7-Db better steady-state performance than the MPNLMS
- MPNLMS Gives improved performance then PNLMS
- PNLMS behavior degrades significantly when identifying not-so-sparse echo channels.
- IPNLMS It does not outperform MPNLMS for highly sparse impulse responses

B. *Performance of Algorithms in Perturbations (non-linear echo or noise) Condition:*

- Performance decreases as the level of perturbations increase but that echo cancellation seems to be more robust to nonlinearities than noise with a similar SNR (with the exception of the FBLMS algorithm).
- The perturbation performance decreases for all adaptive filters.
- As the level of perturbations increase, performance decreases in both non-linear and noisy environments.

## V. CONCLUSION

This paper gives the frame work for developing Acoustic Echo Cancellation Generalized Postfilter in Noisy Environment for Speech Recognition System.

An adaptive filter is used to generate an estimate of the echo signal that is then subtracted from the microphone signal.

Postfilters are commonly used to obtain further echo attenuation.

We have further explained the parameters, affect performance of the algorithms. Performance evaluation of the postfilter algorithm with respective to Residual Echo Suppression, De-reverberation Performance Evaluation and Joint Suppression is explained. Some echo cancellers (filters and algorithms) and their comparative study is discussed.